\documentclass[twocolumn,amssymb, nobibnotes,aps, prb]{revtex4-1}
\usepackage{graphicx}
\usepackage{textcomp}
\usepackage{amsmath}
\usepackage{commath}
\usepackage{color}
	
\begin{document}
	\title{Dynamics of interacting particle systems: Modeling implications of the repulsive interactions and experiments on magnetic prototypes}
	\author{Weijian Jiao}
	\email{jiaox085@umn.edu}
	\author{Stefano Gonella}
	\email{sgonella@umn.edu}
	\affiliation{Department of Civil, Environmental, and Geo- Engineering\\University of Minnesota, Minneapolis, MN 55455, USA\\}
	
	\begin{abstract}
		In this work, we investigate the dynamics of interacting particle systems subjected to repulsive forces, such as lattices of magnetized particles. To this end, we first develop a general model capable of capturing the complete dynamical behavior of interacting particle systems governed by arbitrary potentials. The model elucidates the important role played by the static repulsive forces exchanged between particles in the initial equilibrium configuration, which is distilled and mathematically captured by a dedicated component of the stiffness matrix. The implications of the model are then examined through the simple illustrative example of a magnetic particle oscillator, by which we show that the effect associated with the initial static forces is germane to two- or higher-dimensional particle systems and vanishes for 1D chains. In the context of wave propagation, we show that this type of effect manifests as  modal-selective corrections of the dispersion relation of 2D repulsive lattices. To corroborate these findings, we perform laser vibrometry experiments on a lattice prototype consisting of a triangular grid of magnets supported by an elastic foundation of thin pillars. The tests unequivocally confirm the emergence of distinctive dispersive regimes in quantitative accordance to the model.
		\vspace{0.4cm}
	\end{abstract}
	
\maketitle
\section{Introduction}

Lattices of interacting particles have been used as a versatile modeling proxy to describe many physical and mechanical systems. For example, a gas of electrons interacting with long-range Coulomb forces can crystallize into ordered lattices, known as Wigner crystals \cite{Wigner_1934,Bonsall_1977,Schulz_1993}. At a higher scale, charged microparticle systems, including colloidal dispersions and dusty plasmas, can be origanized, through a Yukawa or a screened-Coulomb interaction, into ordered spatial structures referred to as Yukawa lattices \cite{Pieranski_1980,Peeters_1987, Wang_2001,Nunomura_2002,Kalman_2004}. Crystalline particle ensembles can also be formed at the macroscopic level. A classical example is offered by constrained granular systems composed of beads interacting through Hertzian contacts \cite{Caroli_2002,Coste_2008,Boechler_2010, Tournat_2013, Ganesh_PRL_2015}, possibly under the confining action of compressive loads. Recently, conceptually similar implementations have been obtained using arrays of repulsive magnets \cite{Miguel_2014, Mehrem_2017, Taberlet_2018, Moler_n_2019}. Although the above-mentioned examples are drawn from different physical domains, their dynamical properties are controlled by analogous laws and therefore captured by similar analytical models. One important feature shared across all these problems is that, in these systems, each particle is in equilibrium at rest under the action of self-balancing static forces exchanged with its neighbors. Our objective is to pinpoint through modeling, and verify via experiments, the signature of the static forces on the dynamical behavior of these systems.
	
In conventional treatments of particle systems under small perturbation, it is standard practice to define a potential and expand it in Taylor series ahead of deriving the system's stiffness matrix \cite{Maradudin_1963, Bonsall_1977, dove_1993}. Alternatively, we choose to derive the governing equations starting directly from an arbitrary potential and preserving all the kinematic contributions throughout the derivation of the internal forces and the determination of the stiffness matrix. The procedure yields a stiffness term that is directly linked to the existence of the initial self-balancing static forces and proportional to their strength. To demonstrate the validity of this result and illustrate its potential for capturing the underlying physics of this class of systems, we first apply the model to the benchmark problem of a magnetic oscillator. Here, we show how the incorporation (or lack thereof) of this dedicate stiffness term manifests as a sharp modification of the oscillatory characteristics (e.g, natural frequencies) of the oscillator. We then shift our focus on the propagation of waves in 2D magnetically repulsive lattices, in which the incorporation of the aforementioned stiffness term induces macroscopic dispersion shifts that are heavily wavevector-dependent and mode-sensitive. We corroborate our theoretical findings via explicit time-domain numerical simulations as well as using laser vibrometry experiments carried out on a prototype lattice of magnetized particles supported by an elastic foundation of thin pillars. 
	
In Sec.~II, we establish a complete dynamical model for interacting particle systems with arbitrary potential. In Sec.~III, we analyze the illustrative example. In Sec.~IV, we adapt the analysis to capture the distinctive signature of the repulsive interaction on the dispersive characteristics of 2D magnetic particle lattices, and we corroborate the results with numerical simulations and experiments. The significance and potential impacts of this work are summarized in Sec.~V.

\section{A complete dynamical model for arbitrary potential}
 
Consider a particle system with an initial equilibrium configuration in which the particles interact through an arbitrary potential $\phi(r)$. If the system is perturbed from equilibrium, the potential can be expressed as
\begin{align}\label{1}
\begin{split}
\Phi(r)&=\frac{1}{2}\sum_{i,j \ne i} \phi(\norm{\mathbf{r}_{i,j}})=\frac{1}{2}\sum_{i,j \ne i} \phi \left(  \norm{\mathbf{R}_{i,j}+\mathbf{u}_{i,j}}\right) 
\end{split}
\end{align}  
where $\mathbf{R}_{i,j}$ and $\mathbf{r}_{i,j}$ are the position vectors between particle $i$ and particle $j$ in the initial equilibrium and perturbed configurations, respectively, $\mathbf{u}_{i,j}=\mathbf{u}_{j}-\mathbf{u}_{i} $ is the relative displacement between the two particles, and the specific form of $\Phi(r)$ reflects the physics governing the particle interactions in the system at hand. The force exerted on particle $i$ by the other particles is obtained as
\begin{align}\label{3}
\begin{split}
\mathbf{F}_i=-\nabla \Phi_i(r)=-\sum_{j\ne i} \phi_{r}(r_{i,j}) \mathbf{n}_{i,j}
\end{split}
\end{align}  
where $\mathbf{n}_{i,j}=\frac{\mathbf{r}_{i,j}}{r_{i,j}}$ is the unit vector in the direction connecting particles $i$ and $j$. Following a classical linearization procedure, the stiffness matrix $\mathbf{D}$ is derived as\footnote{For small displacements (i.e., $\norm{\mathbf{u}_{i,j}} \ll \norm{\mathbf{R}_{i,j}} $), it is intuitive to treat $\mathbf{n}_{i,j}$ as constants $\mathbf{n}^0_{i,j}=\frac{\mathbf{R}_{i,j}}{R_{i,j}}$, given the fact that $\mathbf{n}_{i,j} \approx \frac{\mathbf{R}_{i,j}}{R_{i,j}}$ for infinitesimal displacements. This intuitive practice, however, results in the overlooking of the stiffness component denoted as $\mathbf{D}^*$ in this article.}

\begin{align}\label{4}
\begin{split}
\mathbf{D}&=\left.{\nabla }_{\mathbf{u}}\sum_{i} \mathbf{F}_{i}\right|_{\mathbf{u}=\mathbf{0}}\\
&=-\sum_{i}\sum_{j\ne i} \left[ \mathbf{n}_{i,j} \otimes \nabla _{\mathbf{u}} \phi_{r}(r_{i,j}) +   \phi_{r}(r_{i,j}) \nabla _{\mathbf{u}}\mathbf{n}_{i,j}\right]_{\mathbf{u}=\mathbf{0}} \\
&=-\sum_{i}\sum_{j\ne i} \phi_{rr}(R_{i,j}) \mathbf{n}^0_{i,j} \otimes  \mathbf{n}^0_{i,j} + \\& -\sum_{i}\sum_{j\ne i} \frac{\phi_{r}(R_{i,j})}{R_{i,j}} \left( \mathbf{I} - \mathbf{n}^0_{i,j} \otimes  \mathbf{n}^0_{i,j} \right) \\
&\equiv \mathbf{D}_0+\mathbf{D}^*
\end{split}
\end{align} 
where $\mathbf{I}$ is the identity matrix, $\mathbf{n}^0_{i,j}=\frac{\mathbf{R}_{i,j}}{R_{i,j}}$ and $\otimes$ denotes the dyadic product. Detailed derivations of Eq.~\ref{3} and Eq.~\ref{4} are given in the Supplemental Material \cite{SI_PRB_2020}. Interestingly, the treatment yields two distinct contributions to the stiffness matrix. In addition to the conventional term $\mathbf{D}_0$, typical of generic particle systems with masses connected by linear springs, we obtain a secondary component, here denoted as $\mathbf{D}^*$, which depends on the first derivative of the potential $\phi(r)$ evaluated at the unperturbed configuration and is therefore associated with the presence of the initial self-balancing static forces between particles. 

The major implications of a model involving matrix $\mathbf{D}^*$ are summarized in the following key points:

(1) Since the formulation in Eqs.~\ref{1}-\ref{4} is independent of any specific assumptions about the geometry and constitutive behavior of the system, the model endowed with $\mathbf{D}^*$ has universal validity, and can be applied to any particle systems governed by arbitrary potentials including, for example, granular phononic crystals at the macroscale. 

(2) $\mathbf{D}^*$ vanishes in configurations with nearest-neighbor interactions in which $\phi_{r}(R_{i,j}) = 0$. This scenario corresponds to conventional spring-mass systems, where each particle only interacts with its immediate neighbors and its potential is initially at its minimum, resulting in no static forces between particles at rest. 

(3) It can  be shown that $\mathbf{D}^*$ naturally vanishes if the particle system is one-dimensional. This results from the fact that, in 1D systems, the unit vectors $\mathbf{n}_{i,j}$ between pairs of particles remain constant and equal to either [1 0] or [-1 0] even during motion, which implies that their differentiations with respect to the displacements $\mathbf{u}_{i,j}$ are identically null, leading to $\mathbf{D}^*=0$. This unique feature indicates that, unlike $\mathbf{D}_0$, which is pervasive to the equations of motion for any systems, the effect captured by $\mathbf{D}^*$ is germane to high-dimensional (2D and 3D) systems. 


(4) It is worth noticing that the stiffness term $\mathbf{D}^*$ is associated with the fact that the orientation landscape of the static forces varies during motion and its differentiation may yield a finite value upon linearization. Consequently, it is a linear contribution whose effect is amplitude-independent, and therefore does not require large excitations to be activated.

\section{Introductory example: 2D magnetic particle oscillator}

In this section, we apply the general model to a simple particle system with magnetic interactions to document the effect of $\mathbf{D}^*$ on the steady-state vibrational response of 2D particle oscillators.

\subsection{Analytical model}

\begin{figure} [!htb]
	\centering
	\includegraphics[scale=0.7]{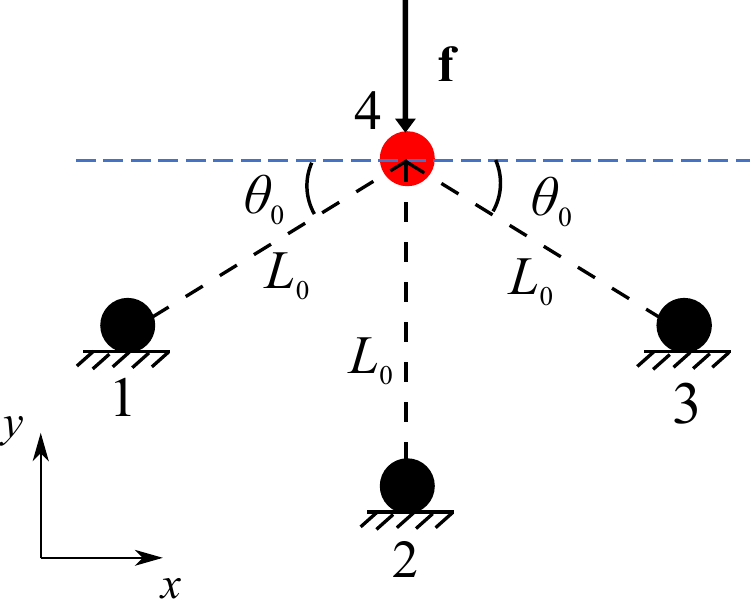}
	\caption{Schematic of a magnetic particle oscillator.}
	\label{2D magnet resonator}
\end{figure}

Consider the system of magnetized particles shown in Fig.~\ref{2D magnet resonator}. Assume that the four particles are identical with mass $M$ and subjected to mutually repulsive forces. To establish the initial equilibrium conditions, a constant vertical force $\mathbf{f}$ is applied on the free particle 4 (red dot) to balance the repulsive forces exerted on 4 by the fixed particles 1, 2, and 3 (black dots). The force between pairs of adjacent particles can be written in the form 
\begin{equation}\label{magnet_force_general}
\mathbf{F}=-\nabla \phi(r)=f(r)\mathbf{n}
\end{equation}
where $\mathbf{n}$ is the unit vector in the direction connecting the particles, and $f(r)=\phi_r(r)$, in general, is taken to obey an inverse power law, i.e., $f(r)=b r^{-a}$.

The governing equation for particle 1 is
\begin{equation}\label{Governing_general_magnet1}
\mathbf{M}\mathbf{\ddot{u}}-\sum_{i=1}^{3} \mathbf{F}_{i}=\mathbf{f}
\end{equation}
where $\mathbf{M}=\begin{bmatrix} M & 0 \\ 0 & M \end{bmatrix}$, $\mathbf{u}=\begin{Bmatrix}u \\ v\\ \end{Bmatrix}$, $\mathbf{F}_{i}=f(r_i) \mathbf{n}_i$, $\mathbf{f}=-\sum_{i=1}^{3} f(L_0)\mathbf{n}_{i}(\mathbf{u}=\mathbf{0})$, and
\begin{equation}\label{coefficient}
\begin{split}
r_i&=\norm{L_0\mathbf{e}_i+\mathbf{u}}
\\
\mathbf{n}_i&=\frac{L_0\mathbf{e}_i+\mathbf{u}}{\norm{L_0\mathbf{e}_i+\mathbf{u}}}
\end{split}
\end{equation}
where $\mathbf{e}_i \thickspace (i=1,2,3)$ can be expressed as $\mathbf{e}_1=\begin{Bmatrix}\cos\theta_0 \\ \sin\theta_0\\ \end{Bmatrix}$, $\mathbf{e}_2=\begin{Bmatrix}0 \\ 1\\ \end{Bmatrix}$, and $\mathbf{e}_3=\begin{Bmatrix}-\cos\theta_0 \\ \sin\theta_0\\ \end{Bmatrix}$.

In the small perturbation limit, the linear stiffness matrix of the system, according to Eq.~\ref{4}, can be obtained as 
\begin{align}\label{stifness_matrix}
\begin{split}
\mathbf{D}&=-\left.{\nabla }_{\mathbf{u}}\sum_{i=1}^{3} \mathbf{F}_{i}\right|_{\mathbf{u}=\mathbf{0}}  \\
&=-\sum_{i=1}^{3} f_{r}(L_0)\mathbf{e}_i\otimes    \mathbf{e}_i  - \sum_{i=1}^{3} \frac{f(L_0)}{L_0}\left( \mathbf{I}-\mathbf{e}_i \otimes \mathbf{e}_i\right) \\
&=-\sum_{i=1}^{3} \phi_{rr}(L_0)\mathbf{e}_i\otimes    \mathbf{e}_i  - \sum_{i=1}^{3} \frac{\phi_{r}(L_0)}{L_0}\left( \mathbf{I}-\mathbf{e}_i \otimes \mathbf{e}_i\right) 
\end{split}
\end{align}
 where the first term is identified as the conventional stiffness matrix $\mathbf{D}_0$, and the second term corresponds to the additional stiffness matrix $\mathbf{D}^*$ introduced in Eq.~\ref{4}.

The natural frequency $\omega_0$ of the system is the only admissible root of the characteristic equation obtained by solving the eigenvalue problem 
\begin{equation}\label{eigenvalue_prob}
\left( -\omega^2 \mathbf{M}+\mathbf{D}_0+\mathbf{D}^*\right) \mathbf{u}=\mathbf{0}
\end{equation}
To quantify the separate contributions of the two stiffness terms, we consider the reference natural frequency $\bar{\omega}_0$ obtained from the conventional model
\begin{equation}\label{eigenvalue_prob_0}
\left( -\omega^2 \mathbf{M}+\mathbf{D}_0\right) \mathbf{u}=\mathbf{0}
\end{equation}
In this example, each particle pair is treated as an ideal magnetic dipole with repulsive force expressed as \cite{Mehrem_2017,Griffiths}
\begin{equation}\label{magnet_force}
\mathbf{F}=\frac{3\mu_0}{4\pi}\frac{m^2}{r^4}\mathbf{n}
\end{equation}
where $\mu_0$ is the permeability of the medium and $m$ is the magnetic moment. Combining Eq.~\ref{stifness_matrix} and Eq.~\ref{magnet_force}, yields
\begin{equation}\label{stifness_matrix_explicit}
\mathbf{D}=\sum_{i=1}^{3} \frac{4\gamma}{L_0^5}\mathbf{e}_i\otimes    \mathbf{e}_i  - \sum_{i=1}^{3} \frac{\gamma}{L_0^5}\left( \mathbf{I}-\mathbf{e}_i \otimes \mathbf{e}_i\right) 
\end{equation}
where $\gamma=\frac{3\mu_0 m^2}{4\pi}$. For an arbitrary choice of parameters ($M=1$, $L_0=1$, $\gamma=10^4$ with standard SI units used throughout the paper except where specified), the natural frequencies can be calculated from Eq.~\ref{eigenvalue_prob} and Eq.~\ref{eigenvalue_prob_0}, and the values will be compared with simulation results provided in section III.~2.

\subsection{Time-domain simulations}

We perform time-domain numerical simulations assuming a harmonic excitation applied vertically on particle 1 (note that, in the simulation, the horizontal motion of the particle is constrained to ensure stability, as explained in the Supplemental Material \cite{SI_PRB_2020}). The governing equation (Eq.~\ref{Governing_general_magnet1}) is integrated in time using the Verlet algorithm \cite{Swope_1982}, and the magnitude of the harmonic response is recorded after steady-state conditions are reached. To effectively establish steady-state conditions, we add viscous damping to the system and we consider sufficiently long excitation times to fully dissipate the signature of the transient response. We also keep the amplitude of excitation sufficiently low to neglect the effects of nonlinearity (naturally embedded in the constitutive model of Eq.~\ref{magnet_force}), which are not relevant for this treatment. 

\begin{figure} [!htb]
	\centering
	\includegraphics[scale=0.32]{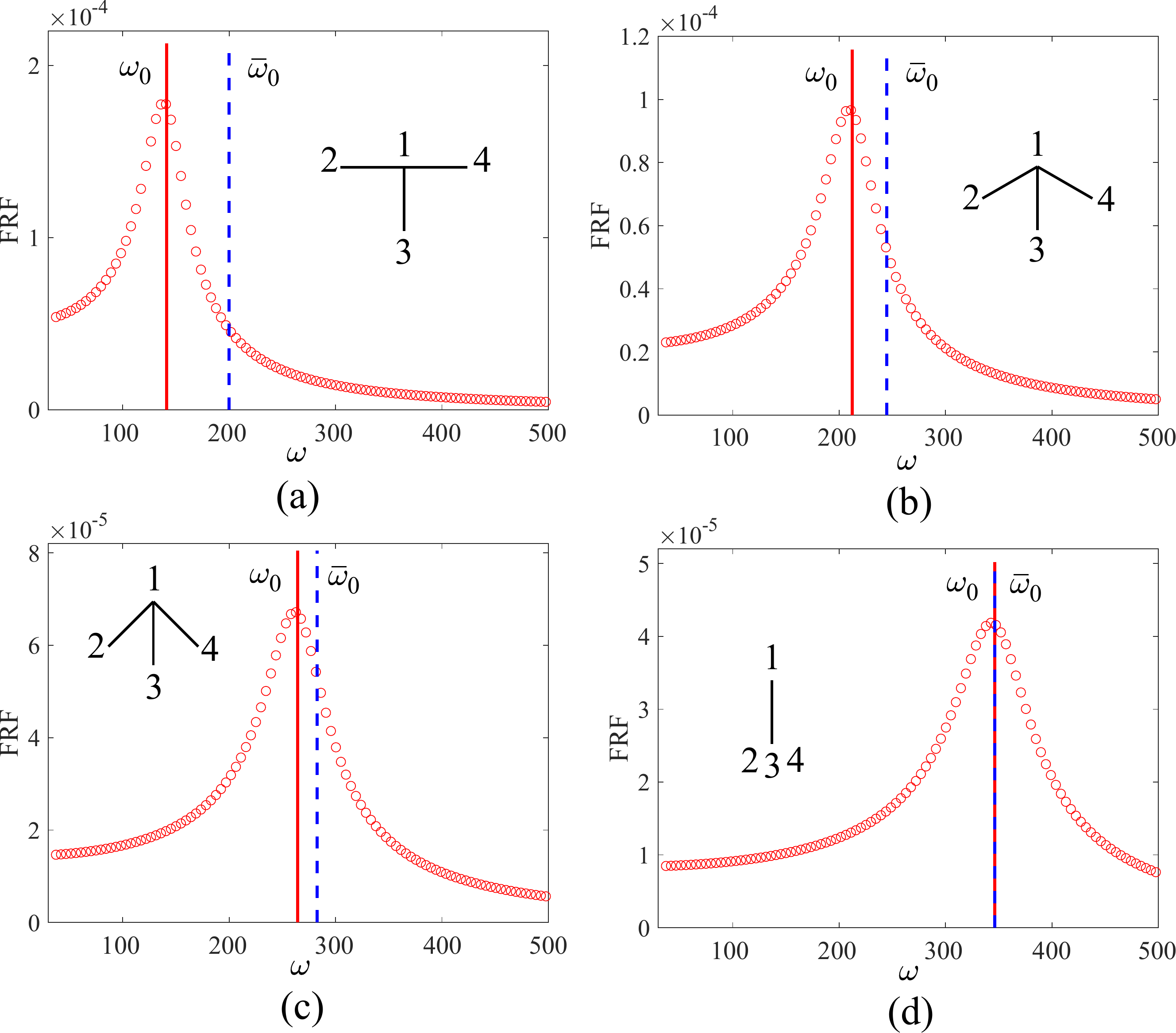}
	\caption{Frequency response function (FRF) of the complete systems for four configurations, sketched in the insets: (a) $\theta_0=0$, (b) $\theta_0=\pi/6$, (c) $\theta_0=\pi/4$, and (d) $\theta_0=\pi/2$. The natural frequencies computed from the analytical models are also reported as vertical lines.}
	\label{2D magnet resonator_FRF}
\end{figure}

In Fig.~\ref{2D magnet resonator_FRF}, we plot the numerically-obtained frequency response function (FRF) of the complete system (red curve with $\textrm{o}$ markers) for four different orientations of the slant links: $\theta_0 \in [0\;\; \pi/6\;\; \pi/4\;\; \pi/2]$. We also superimpose the natural frequency bar $\omega_0$ (red line) and the reference frequency bar $\bar{\omega}_0$ (blue dashed line) predicted from Eq.~\ref{eigenvalue_prob} and Eq.~\ref{eigenvalue_prob_0}, respectively. We observe that the computed natural frequencies match the peaks of the FRF curves. Since the numerical simulations are not subjected to any restrictive assumptions, as they involve updating the most general form of the interaction law at each integration step, this result confirms the validity of the complete model in Eq.~\ref{stifness_matrix}. Moreover, we observe that the difference between $\omega_0$ and $\bar{\omega}_0$ decreases as $\theta_0$ increases. In the limit case of $\theta_0 = \pi/2$, the corresponding system is reduced to a 1D configuration and the two frequencies become identical. This result supports our general remark anticipated earlier that the additional dynamical effect captured by $\mathbf{D}^*$ (here manifests as the frequency difference $\Delta \omega_0 = \omega_0 - \bar{\omega}_0$) is germane to 2D configurations. This marks a fundamental difference with the terms depending on $\mathbf{D}_0$, which do not vanish for 1D configurations. Another interesting feature is that, since the magnetic force is repulsive in our framework, the frequency difference $\Delta \omega_0$ is necessarily negative (softening effect). Finally, we note that, for certain parameter choices, a special condition may occur when $|\Re(\Delta \omega_0)|$ is no longer less than $\bar{\omega}_0$ or when the horizontal motion is not constrained, resulting in dynamical instabilities (a preliminary stability analysis of the magnetic system based on Eq.~\ref{stifness_matrix_explicit} is reported in the Supplemental Material \cite{SI_PRB_2020}). 

\section{Wave propagation in 2D Repulsive Lattices}

In this section, we shift our attention to wave propagation problems. First, we theoretically and numerically investigate the propagation of waves in repulsive lattices of magnetized particles, demonstrating the effect of $\mathbf{D}^*$ on the dispersion relation. We then proceed to experimentally confirm the findings via laser vibrometry experiments.
 
\subsection{Analytical model}
\begin{figure} [!htb]
	\centering
	\includegraphics[scale=0.3]{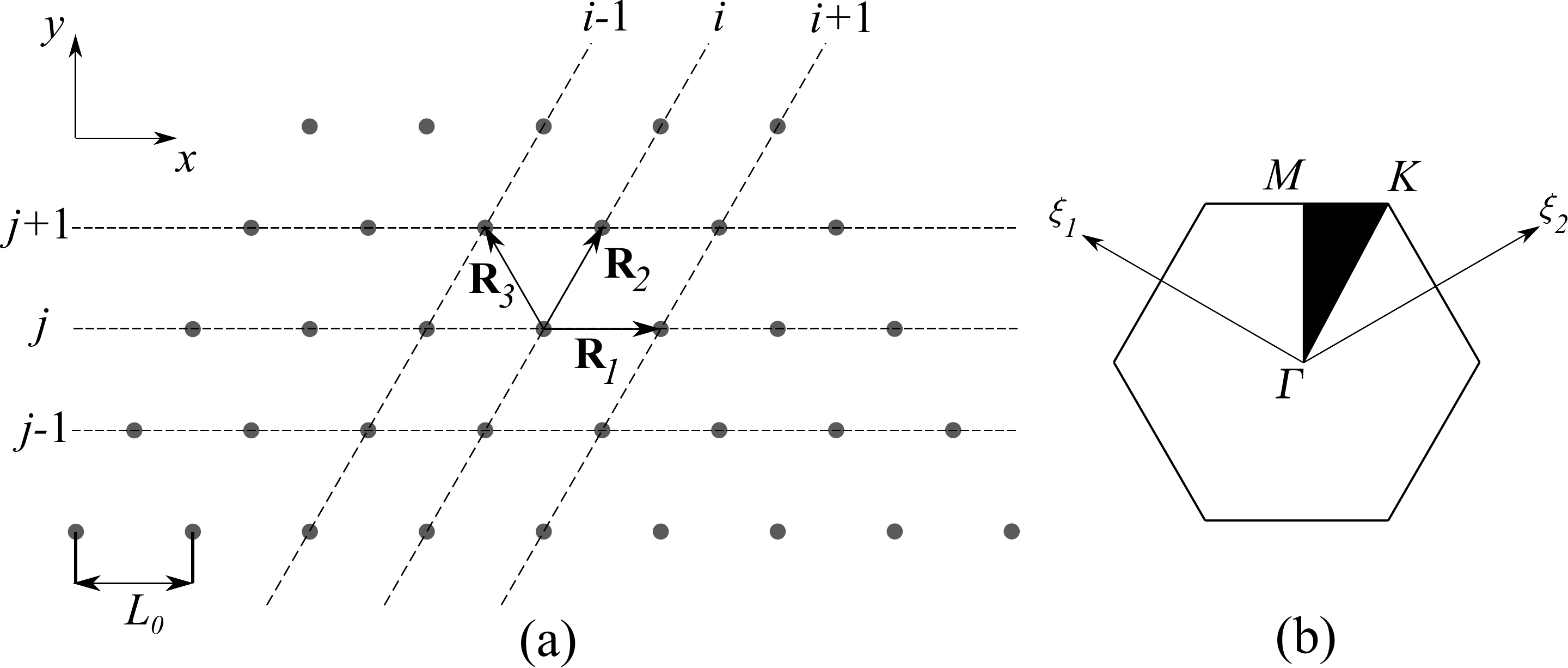}
	\caption{(a) 2D triangular lattice with basis vectors $\mathbf{R}_2$ and $\mathbf{R}_3$. (b) First Brillouin zone with irreducible zone highlighted in black ($\xi_1$ and $\xi_2$ are the components of the nondimentional wavevector in reciprocal space).}
	\label{2D triangular_lattice}
\end{figure}

Consider a triangular lattice consisting of repulsive particles shown in Fig.~\ref{2D triangular_lattice}. For simplicity, each particle in the lattice is assumed to interact only with its nearest neighbors. The governing equation for a particle at location $\mathbf{R}_{i,j}$ can be written as
\begin{equation}\label{Governing_general_triangular_lattice}
\mathbf{M}\mathbf{\ddot{u}}_{i,j}(\mathbf{R}_{i,j},t)+\sum_{l=-3}^{3} \mathbf{F}_{l}(\mathbf{u})=\mathbf{0}
\end{equation}
According to Eq.~\ref{magnet_force_general}, the repulsive force $\mathbf{F}_{l}$ between two adjacent particles takes the general form
\begin{equation}\label{Governing_general_Force}
\mathbf{F}_l(\mathbf{u})=
\begin{cases}
f(\norm{ \Delta \mathbf{u}_{\pm 1,0}\pm\mathbf{R}_1})\frac{\Delta \mathbf{u}_{\pm 1,0}\pm \mathbf{R}_1}{\norm{ \Delta \mathbf{u}_{\pm 1,0}\pm \mathbf{R}_1}} \\ \mathrm{where} \; \Delta \mathbf{u}_{\pm 1,0}=\mathbf{u}_{i\pm 1,j}-\mathbf{u}_{i,j}, & \text{if}\ l=\pm 1 \\
f(\norm{\Delta \mathbf{u}_{0,\pm 1}\pm \mathbf{R}_2})\frac{\Delta \mathbf{u}_{0,\pm 1}\pm \mathbf{R}_2}{\norm{\Delta \mathbf{u}_{0,\pm 1}\pm \mathbf{R}_2}}\\ \mathrm{where} \; \Delta \mathbf{u}_{0,\pm 1}=\mathbf{u}_{i,j\pm 1}-\mathbf{u}_{i,j}, & \text{if}\ l=\pm 2 \\
f(\norm{\Delta \mathbf{u}_{\mp 1,\pm 1}\pm \mathbf{R}_3})\frac{\Delta \mathbf{u}_{\mp 1,\pm 1}\pm \mathbf{R}_3}{\norm{\Delta \mathbf{u}_{\mp 1,\pm 1}\pm \mathbf{R}_3}}\\ \mathrm{where} \; \Delta \mathbf{u}_{\mp 1,\pm 1}=\mathbf{u}_{i\mp 1,j\pm 1}-\mathbf{u}_{i,j}, & \text{if}\ l=\pm 3
\end{cases}
\end{equation}

\vspace{0.1cm}
\noindent where $\mathbf{R}_1=L_0\mathbf{e}_1=L_0\begin{Bmatrix}1 \\ 0\\ \end{Bmatrix}$, $\mathbf{R}_2=L_0\mathbf{e}_2=L_0\begin{Bmatrix}1/2 \\ \sqrt{3}/2\\ \end{Bmatrix}$, and $\mathbf{R}_3=L_0\mathbf{e}_3=L_0\begin{Bmatrix}-1/2 \\ \sqrt{3}/2\\ \end{Bmatrix}$.
Assuming that the displacement $\Delta \mathbf{u}$ is infinitesimally small, Eq.~\ref{Governing_general_triangular_lattice} can be linearized as 
\begin{multline}\label{Governing_general_triangular_linearized}
\mathbf{M}\mathbf{\ddot{u}}_{i,j}+\sum_{l=1}^{3}\left\lbrace \left[  f_r(L_0)\mathbf{e}_l\otimes    \mathbf{e}_l\right]\Delta \mathbf{u}_l \right\rbrace \\  + \sum_{l=1}^{3}\left\lbrace \left[\frac{f(L_0)}{L_0}\left( \mathbf{I}-\mathbf{e}_l \otimes \mathbf{e}_l\right) 
\right]\Delta \mathbf{u}_l \right\rbrace =\mathbf{0}
\end{multline}
where $\Delta \mathbf{u}_l= 
\begin{cases}
\mathbf{u}_{i+1,j}+\mathbf{u}_{i-1,j}-\mathbf{u}_{i,j}, & \text{if}\ l= 1 \\
\mathbf{u}_{i,j+1}+\mathbf{u}_{i,j-1}-\mathbf{u}_{i,j}, & \text{if}\ l=2 \\
\mathbf{u}_{i-1,j+1}+\mathbf{u}_{i+1,j-1}-\mathbf{u}_{i,j}, & \text{if}\ l= 3
\end{cases}$.

A plane wave solution of Eq.~\ref{Governing_general_triangular_linearized}, with wave vector $\boldsymbol{k}$ and frequency $\omega$, is given as
\begin{equation} \label{soln_plane_wave}
\mathbf{u}_{i,j}=A \boldsymbol{\phi} e^{i(\boldsymbol{k} \cdot \mathbf{R}_{i,j}+\omega t)}
\end{equation} 
where  $A$ is a constant, and $\boldsymbol{\phi}=\begin{Bmatrix} \phi_u \\ \phi_v\\ \end{Bmatrix}$ is a modal vector. 
According to Floquet-Bloch theorem, the relations between displacements at neighboring sites can be expressed as 
\begin{equation}\label{Bloch_Condition}
\mathbf{u}_{i\pm 1,j\pm 1}=\mathbf{u}_{i,j}e^{i(\pm \boldsymbol{k} \cdot \mathbf{R}_1 \pm \boldsymbol{k} \cdot \mathbf{R}_2)}
\end{equation}
Substituting Eq.~\ref{soln_plane_wave} and Eq.~\ref{Bloch_Condition} into Eq.~\ref{Governing_general_triangular_linearized}, yields the wavevector-dependent eigenvalue problem
\begin{equation}\label{Eigenvalue_prob_triangular}
\left[ -\omega^2\mathbf{M} +\mathbf{D}(\boldsymbol{k})\right] \boldsymbol{\phi}=\mathbf{0}
\end{equation}
where
\begin{multline}\label{Stiffness_matrix}
\mathbf{D}(\boldsymbol{k})=2\sum_{l=1}^{3}\left\lbrace  f_r(L_0)\mathbf{e}_l\otimes    \mathbf{e}_l\left[ \cos(\boldsymbol{k} \cdot \mathbf{R}_l)-1\right]  \right\rbrace \\ + 2\sum_{l=1}^{3}\left\lbrace\frac{f(L_0)}{L_0}\left( \mathbf{I}-\mathbf{e}_l \otimes \mathbf{e}_l\right) 
\left[ \cos(\boldsymbol{k} \cdot \mathbf{R}_l)-1\right]  \right\rbrace
\end{multline}
is a wavevector-dependent stiffness matrix. Again, we observe the appearance of two terms in the stiffness matrix. Canonically, the linear dispersion relation of the magnetic system is obtained by solving the eigenvalue problem for wavevectors along the contour of the irreducible Brillouin zone. Based on the conventional stiffness matrix, we can also define a reference system, whose dispersion relation is obtained from the following eigenvalue problem
\begin{equation}\label{Eigenvalue_prob_triangular_reference}
\left[ -\omega^2\mathbf{M} +\mathbf{D}_0(\boldsymbol{k})\right] \boldsymbol{\phi}=\mathbf{0}
\end{equation}
where $\mathbf{D}_0(\boldsymbol{k})$ is the first term of $\mathbf{D}(\boldsymbol{k})$ in Eq.~\ref{Stiffness_matrix}. 

\begin{figure} [!htb]
	\centering
	\includegraphics[scale=0.5]{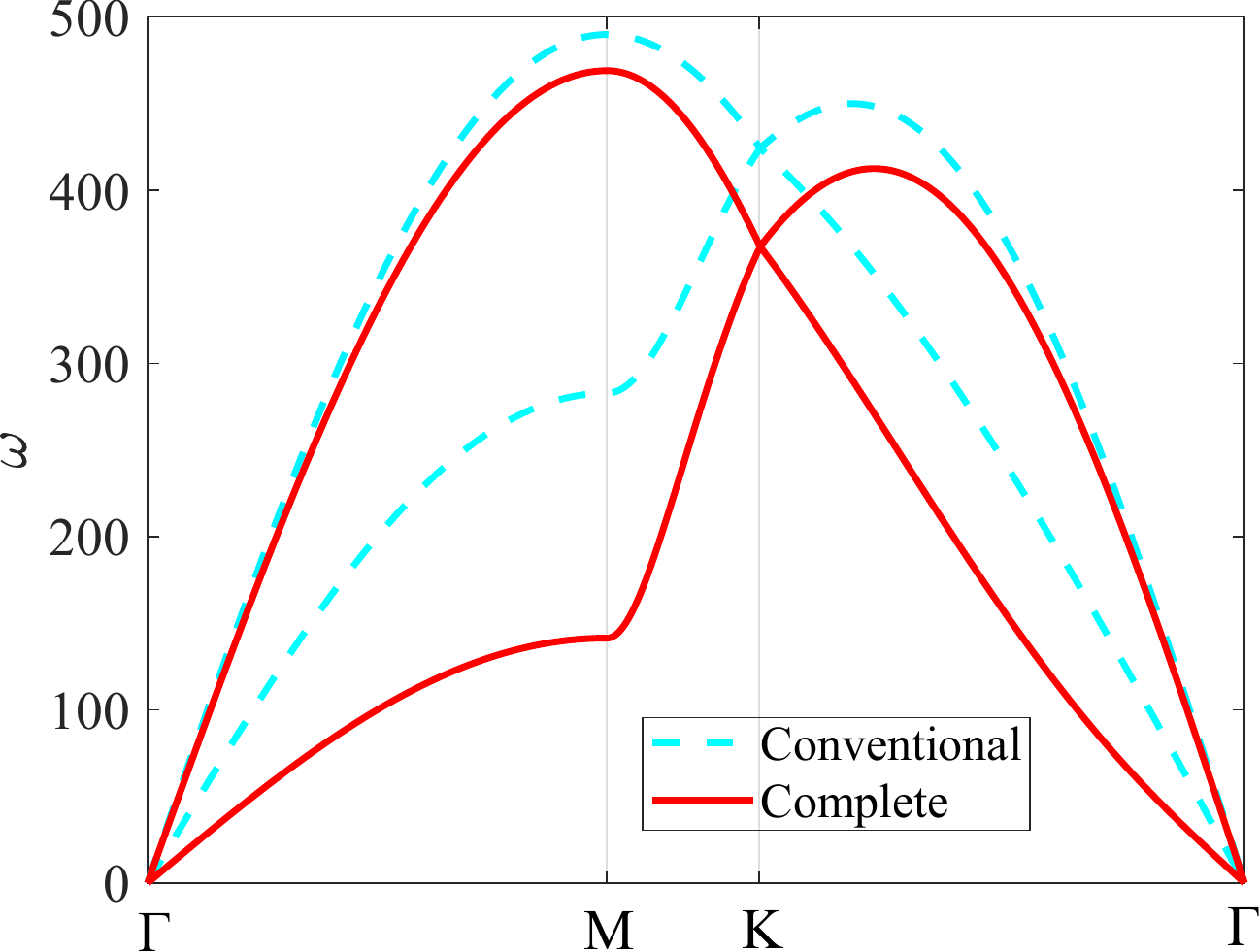}
	\caption{Dispersion relations of the repulsive lattice obtained from the complete and conventional (reference) models.}
	\label{Band_diagram_triangular}
\end{figure}

In Fig.~\ref{Band_diagram_triangular}, we plot the band diagram (red curves) of the 2D triangular repulsive lattice with magnetic interaction law (according to Eq.~\ref{magnet_force}), and we superimpose the reference one (light blue dashed curves) obtained solving Eq.~\ref{Eigenvalue_prob_triangular_reference}. Clearly, $\mathbf{D}^*(\boldsymbol{k})$ (i.e., the second term in Eq.~\ref{Stiffness_matrix}) has significant influence on the dispersion relation of the repulsive lattice, especially for the first band.

\subsection{Full-scale simulations}

\begin{figure} [!htb]
	\centering
	\includegraphics[scale=0.6]{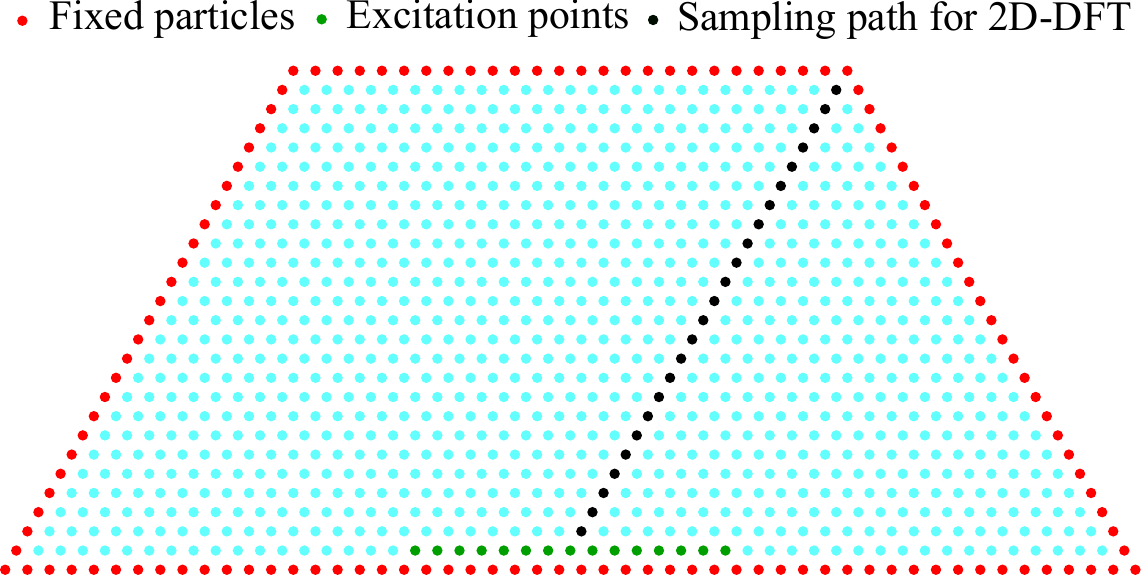}
	\caption{Finite lattice used in the full-scale simulations.}
	\label{Systematic_Triangular_Lattice_Simulation}
\end{figure}

To validate our analytical model, we perform a suite of full-scale simulations to obtain the wave response of a finite lattice (shown in Fig.~\ref{Systematic_Triangular_Lattice_Simulation}) and we compare it against that of a corresponding reference system. The particles located on the boundary (red dots) are fixed in order to establish initial equilibrium conditions. The objective is to numerically reconstruct the band diagram and compare it against that of a corresponding reference system. We consider wavevectors $\boldsymbol{k}$ sampled along the $\mathrm{\Gamma}$-$\mathrm{M}$ direction, which correspond to wave propagation in the vertical direction. Nearly plane-wave conditions are established by considering an array of excitation points collocated at the particles denoted as green dots in Fig.~\ref{Systematic_Triangular_Lattice_Simulation}. The force excitation is prescribed as a five-cycle tone burst with carrier frequency $\Omega_0$ chosen to fall within the frequency range of the first (shear-dominant) and of the second (longitudinal) modes, respectively. A small-amplitude force is applied in the horizontal (vertical) direction to optimally excite the shear (longitudinal) mode in the linear regime.

\begin{figure} [!htb]
	\centering
	\includegraphics[scale=0.31]{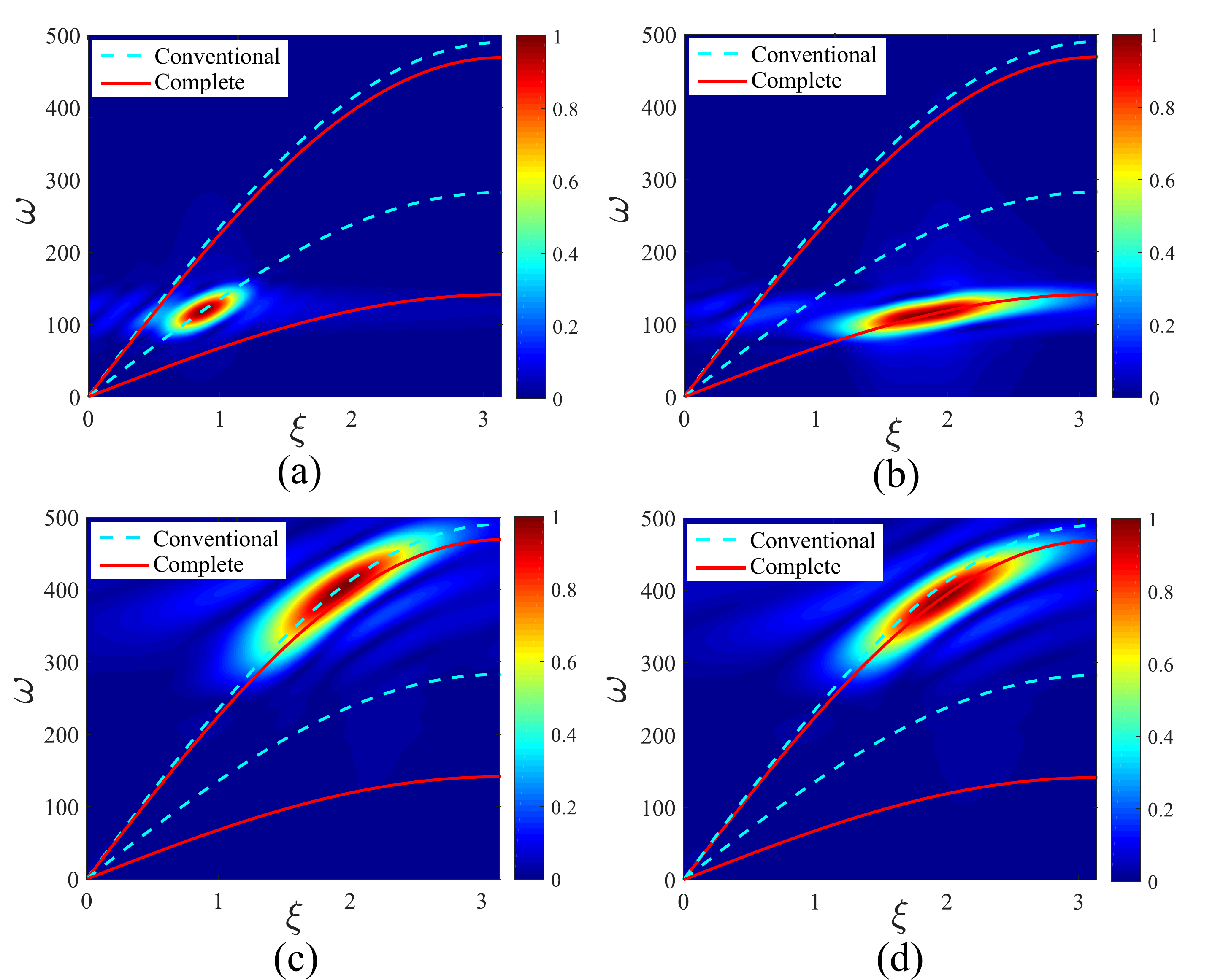}
	\caption{Spectral response from numerical simulations compared against band diagrams from Bloch analysis . (a) and (b) Response to excitation at $\Omega_0=120$ rad/s. (c) and (d) Response to excitation at $\Omega_0=400$ rad/s. (a) and (c) Response of the conventional (reference) system; (b) and (d) Response of the complete system.}
	\label{Dispersion_shift_spectrum}
\end{figure}

The spatio-temporal displacement response is sampled at nodes located along lattice vector $\mathbf{R}_2$ (black dots in Fig.~\ref{Systematic_Triangular_Lattice_Simulation}) and transformed via 2D Discrete Fourier Transform (2D-DFT). We perform and compare two simulations. In the first, which we refer to as``complete simulation'', we update at each time step the most general form of the internal force (both its magnitude and direction) based on Eq. 14. The second is a conventional small-amplitude linear simulation, in which only the magnitude is updated at each step.

For an excitation at $\Omega_0=120$ rad/s (in the shear mode range), the normalized spectral amplitude maps obtained from the complete and conventional simulations are plotted in Fig.~\ref{Dispersion_shift_spectrum}(a-b), respectively. For excitation at $\Omega_0=400$ rad/s (in the longitudinal mode range), the results are plotted in Fig.~\ref{Dispersion_shift_spectrum}(c-d). From a visual inspection, we observe that the results from the complete simulation match the dispersion relation predicted using the complete analytical model (i.e., Eq.~\ref{Eigenvalue_prob_triangular}), while the conventional simulation results in large dispersion deviations for the shear mode and minor ones for the longitudinal mode. 
\footnote{However, it shows a good agreement with the conventional spring-mass model, which implies that the overlooking of $\mathbf{D}^*$ is ``forgiving'' if conventional treatments are implemented in both numerical and analytical models.}  These results numerically confirm the inclusion (or lack thereof) of $\mathbf{D}^*$ bears non-negligible modal-selective effects on the prediction of the dispersive behavior of repulsive lattices.

\subsection{Experiments}

\begin{figure} [!htb]
	\centering
	\includegraphics[scale=0.06]{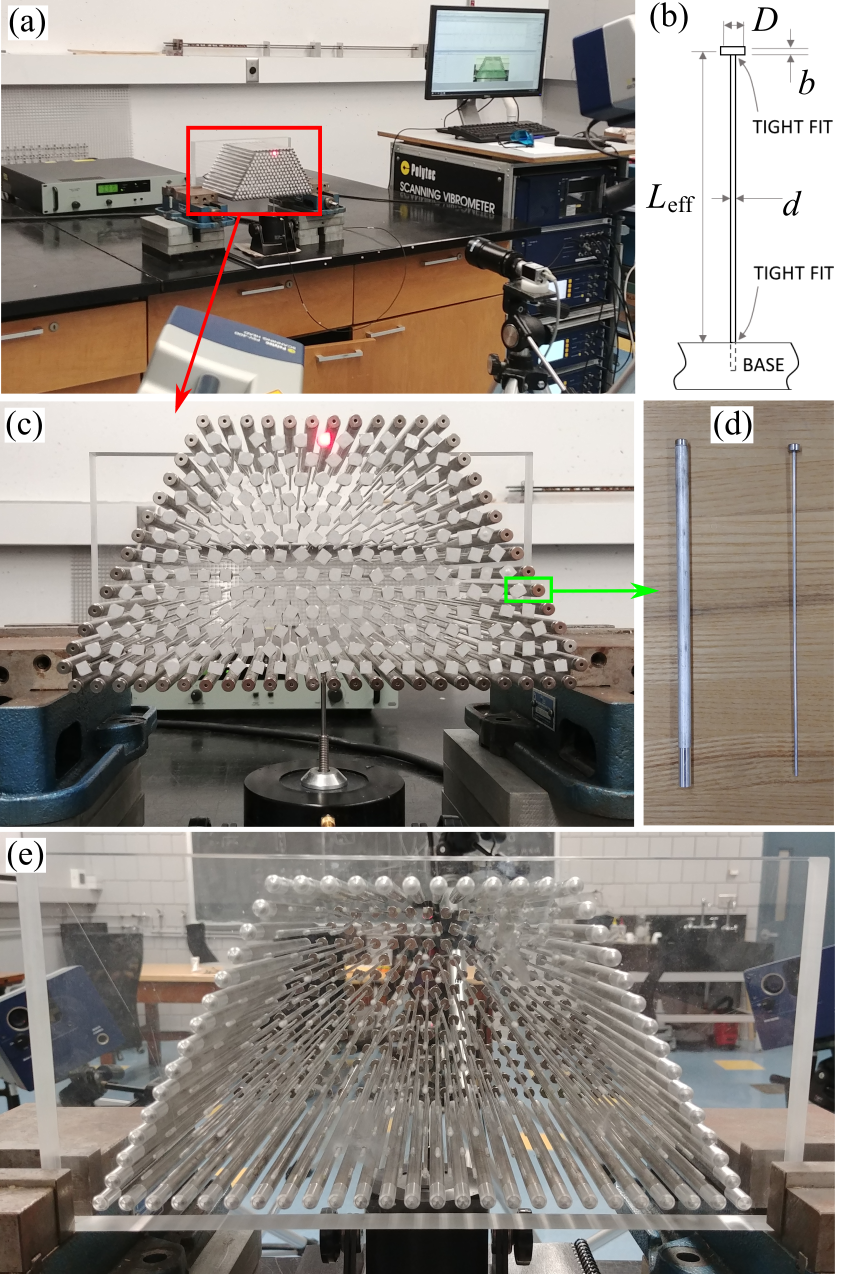}
	\caption{(a) Experimental setup. (b) Schematic of one interior beam-magnet unit plugged into the base. (c) Front view of the magnetic lattice specimen highlighting the shaker position for in-plane excitation. The interior magnets are covered by reflective tape to enhance laser measurements. (d) Magnet-beam units used at boundaries (left) and in the interior of the triangular lattice (right) (e) Back view through the transparent Acrylic base.}
	\label{Experimental_setup}
\end{figure}

To corroborate the theoretical predictions, and to justify the need of the complete model to capture the proper dynamics of realistic repulsive particle systems, we perform a series of experiments on a lattice prototype, shown in Fig.~\ref{Experimental_setup}, which involves finite-size magnets supported by simple structural elements and represents a practical implementation of the idealized system considered in our model. The role of the particles is played by small ring magnets (with outer diameter $D=1/4$ inch $\times$ inner diameter $d=1/16$ inch $\times$ thickness $b=1/8$ inch, Grade N42) interacting repulsively in their own plane and thus spontaneously occupying the nodal locations of a triangular lattice at equilibrium. To enforce planarity of the lattice, the magnets are supported by Aluminum cantilever beams tightly inserted in the magnets ring holes at one tip and clamped to an Acrylic base through a lattice of drilled holes
(see schematic in Fig.~\ref{Experimental_setup}(b)). The interior magnets are supported by slender(with cross sectional diameter 1/16 inch) and highly flexible beams that allow minimally impeded in-plane displacement of the tip magnets, while the exterior magnets, located along the perimeter of a half hexagon, are supported by thick beams (with cross sectional diameter 1/4 inch) featuring large bending stiffness to establish fixed boundary conditions (Fig.~\ref{Experimental_setup}(c-d)). To properly incorporate the effects of the supporting beams, which effectively act as an elastic foundation, into the model used for our reference Bloch analysis, we need to modify the unit cell configuration. To this end, we endow each particle with an additional flexural spring connecting the particle to a fixed ground, here representative of the acrylic base. The spring features an elastic constant proportional to the equivalent bending stiffness of a thin cantilever, which depends on the material and cross-sectional properties of the cantilever beam and on its effective length $L_{\mathrm{eff}}$ (here equal to 15 cm). In order to capture the precise values of the repulsive forces between magnet pairs exhibited by the specific set of magnets used in our test, we perform a static experimental characterization of the magnet-magnet interaction. Details on the modified model as well as the characterization of the magnets are reported in the Supplemental Material \cite{SI_PRB_2020}.

\begin{figure} [!htb]
	\centering
	\includegraphics[scale=0.31]{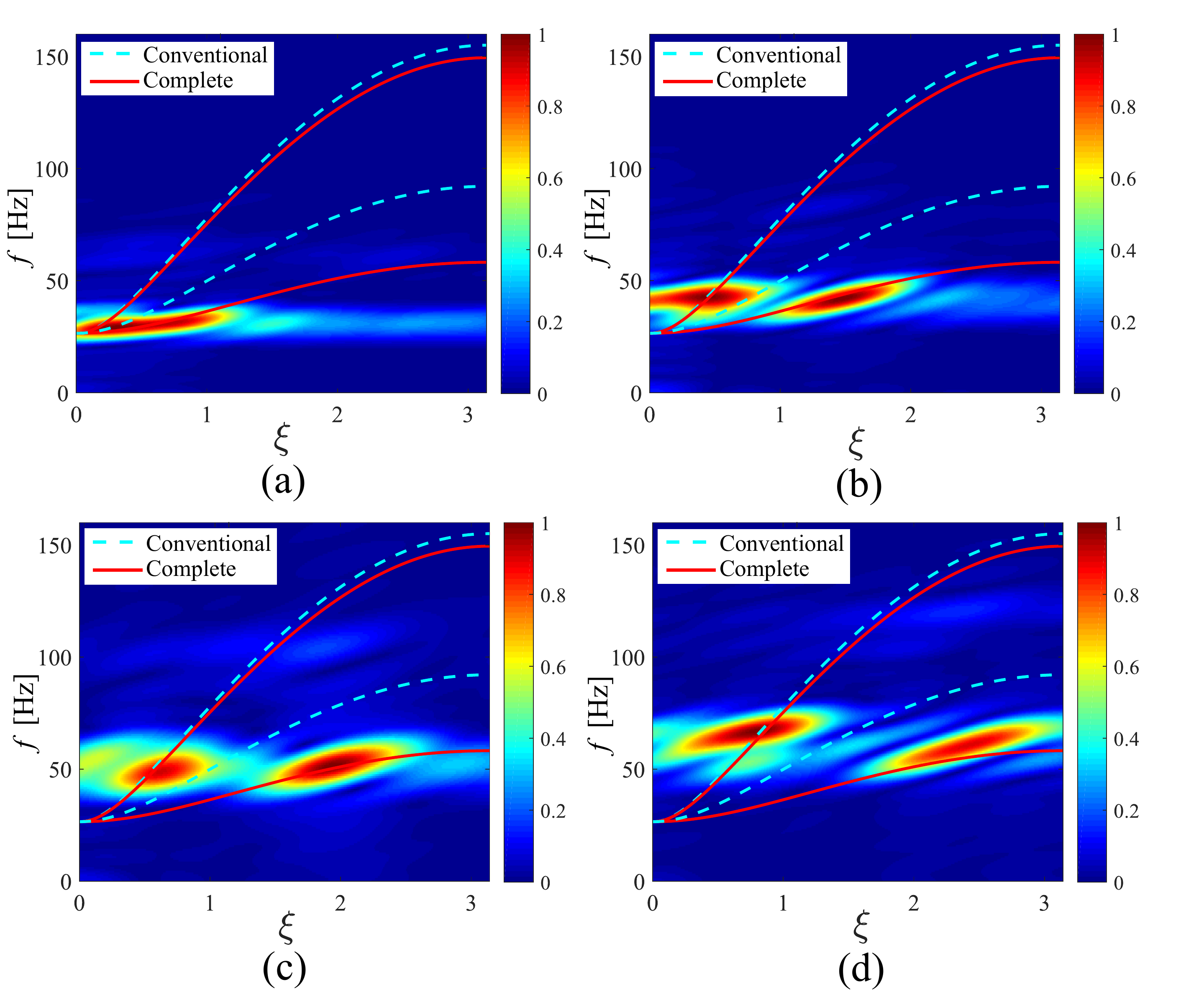}
	\caption{Experimental response spectra for tone-burst excitations at (a) $30$ Hz, (b) $40$ Hz, (c) $50$ Hz, and (d) $60$ Hz. The amplitude spectra conform to the band diagram predicted using the complete model, while they are not properly captured by the conventional one (especially for the shear mode).}
	\label{Spectrum_Experiment}
\end{figure}

The experimental setup is shown in Fig.~\ref{Experimental_setup}(a). A 3D scanning laser Doppler vibrometer (SLDV, Polytec PSV-400-3D) is used to scan the magnets and measure their in-plane response. The excitation is prescribed in the vertical direction at the magnet located at the center of the bottom edge (one layer insider the fixed boundary) through a Bruel \& Kjaer Type $4809$ shaker (powered by a Bruel \& Kjaer Type $2718$ amplifier), as shown in Fig.~\ref{Experimental_setup}(c). In Fig.~\ref{Spectrum_Experiment}, we plot the spectral response obtained via 2D-DFT of the experimental spatio-temporal data sampled along a lattice vector for tone-burst excitations with carrier frequencies centered at 30 Hz, 40 Hz, 50 Hz, and 60 Hz, respectively. For comparison, we superimpose the dispersion relations predicted using our modified analytical model (encompassing both $\mathbf{D}_0$ and $\mathbf{D}^*$ contributions as per Eq.~\ref{Stiffness_matrix}). In contrast with the previous cases, the band diagram is here fully gapped at low frequencies, which is a typical feature of systems with elastic foundations. Notwithstanding small deviations at higher frequencies (which can be easily attributed to non-idealities and unavoidable minor differences between particle model and physical specimen, e.g., the neglecting of possible mild long-range interactions in the theoretical model), the experimental results show remarkable agreement with the dispersion branches obtained from the complete model. This result provides unequivocal experimental evidence supporting the notion that $\mathbf{D}^*$ exerts a profound effect on the dispersion relation. Moreover, the experimental spectra confirm that the effect of $\mathbf{D}^*$ in the shear mode are indeed much stronger than those observed for the longitudinal mode, which is another peculiar characteristic resulting from $\mathbf{D}^*$.
\vspace{0.5cm}
\section{Conclusions}

In this study, we have predicted theoretically and demonstrated experimentally the complete dynamics of 2D interacting particle systems. First, we have formulated a complete theoretical model for particle systems, revealing the existence of a special contribution (denoted as $\mathbf{D}^*$) to the stiffness matrix. Through an illustrative example of a resonating particle system, we have shown that the effect captured by $\mathbf{D}^*$ is intrinsically tied to the 2D nature of the particle arrangements and disappears when the system reduces to a 1D configuration. Then, we have discussed the implications of this effect on the wave propagation characteristics of repulsive lattices, and we have highlighted that the existence of $\mathbf{D}^*$ is responsible for the emergence of mode-selective dispersion shifts. Finally, we have experimentally demonstrated these findings with a lattice prototype assembled using magnets supported by a foundation of beam elements. Besides the magnetic systems, we believe that the general framework presented in this work is applicable to a broad class of physical systems in which particles are subjected to repulsive interactions.

\section*{Acknowledgement}
This work is supported by the National Science Foundation (CAREER Award CMMI-$1452488$). The authors are indebted to Lijuan Yu for her help with the specimen assembly and to Joseph Labuz, Xiaoran Wang and Chen Hu for sharing their invaluable expertise with the force testing apparatus. 
	\bibliography{myrefs}
	
\clearpage
\onecolumngrid

\setcounter{figure}{0}
\setcounter{equation}{0}
\setcounter{page}{1}
\renewcommand{\thefigure}{S\arabic{figure}}
\renewcommand{\theequation}{S\arabic{equation}}
\renewcommand{\thepage}{S\arabic{page}}
		\section*{\Large Supplemental Material}
		\subsection{Derivation of the internal force}
		The force exerted on particle $i$ (i.e., Eq.~2 in the article) is derived as
		\begin{align}\label{S1}
		\begin{split}
		\mathbf{F}_i&=-\nabla \Phi_i(r) \\
		&=-\sum_{j\ne i}\sum_{\alpha} \phi_{\alpha}(r_{i,j})\mathbf{e}^\alpha \\
		&=-\sum_{j\ne i}\sum_{\alpha} \phi_{r}(r_{i,j})\frac{\partial r_{i,j}}{\partial \alpha }\mathbf{e}^\alpha \\
		&=-\sum_{j\ne i}\sum_{\alpha} \phi_{r}(r_{i,j})\frac{R^\alpha_{i,j}+u^\alpha_{i,j}}{r_{i,j}}\mathbf{e}^\alpha\\
		&=-\sum_{j\ne i} \phi_{r}(r_{i,j}) \mathbf{n}_{i,j}
		\end{split}
		\end{align} 
		where $\mathbf{e}^\alpha$ are the Cartesian unit vectors and $r_{i,j}=\sqrt{(R^x_{i,j}+u^x_{i,j})^2+(R^y_{i,j}+u^y_{i,j})^2}$.
		
		\subsection{Derivation of the complete stiffness matrix}
		The complete stiffness matrix $\mathbf{D}$ for lattices of particles interacting through an arbitrary potential $\phi(r)$ (i.e., Eq.~3 in the article) is derived as follows
		\begin{align}\label{S2}
		\begin{split}
		\mathbf{D}&=\left.{\nabla }_{\mathbf{u}}\sum_{i} \mathbf{F}_{i}\right|_{\mathbf{u}=\mathbf{0}} = -\left.{\nabla }_{\mathbf{u}}\sum_{i} \sum_{j\ne i} \phi_{r}(r_{i,j}) \mathbf{n}_{i,j}\right|_{\mathbf{u}=\mathbf{0}}\\
		&=-\sum_{i}\sum_{j\ne i} \left[ \mathbf{n}_{i,j} \otimes \nabla _{\mathbf{u}} \phi_{r}(r_{i,j}) +   \phi_{r}(r_{i,j}) \nabla _{\mathbf{u}}\mathbf{n}_{i,j}\right]_{\mathbf{u}=\mathbf{0}} \\
		&=-\sum_{i}\sum_{j\ne i}\left[  \phi_{rr}(r_{i,j}) \mathbf{n}_{i,j}\otimes \mathbf{n}_{i,j} \right]_{\mathbf{u}=\mathbf{0}} -\sum_{i}\sum_{j\ne i} \phi_{r}(r)\left[ \frac{\mathbf{I}}{r_{i,j}}-\frac{\left(\mathbf{R}_{i,j}+\mathbf{u}_{i,j}\right)\otimes \left(\mathbf{R}_{i,j}+\mathbf{u}_{i,j}\right)}{r_{i,j}^3}\right]_{\mathbf{u}=\mathbf{0}} \\
		&=-\sum_{i}\sum_{j\ne i} \phi_{rr}(R_{i,j}) \mathbf{n}^0_{i,j} \otimes  \mathbf{n}^0_{i,j} -\sum_{i}\sum_{j\ne i} \frac{\phi_{r}(R_{i,j})}{R_{i,j}} \left( \mathbf{I} - \mathbf{n}^0_{i,j} \otimes  \mathbf{n}^0_{i,j} \right) \\
		&\equiv \mathbf{D}_0+\mathbf{D}^*
		\end{split}
		\end{align} 
		in which $\nabla _{\mathbf{u}} \phi_{r}(r_{i,j})= \phi_{rr}(r_{i,j}) \mathbf{n}_{i,j}$ obtained following a procedure similar to what presented in Eq.~\ref{S1}, and  $\nabla _{\mathbf{u}}\mathbf{n}_{i,j} = \begin{bmatrix} \partial n^x_{i,j}/ \partial u^x_{i,j} & \partial n^x_{i,j}/ \partial u^y_{i,j} \\ \partial n^y_{i,j}/ \partial u^x_{i,j} & \partial n^y_{i,j}/ \partial u^y_{i,j} \end{bmatrix}$, whose components are derived as 
		\begin{align}\label{S3}
		\begin{split}
		\partial n^x_{i,j}/ \partial u^x_{i,j}&= \partial( \frac{R^x_{i,j}+u^x_{i,j}}{r_{i,j}})/ \partial u^x_{i,j} = \frac{1}{r_{i,j}}-\frac{(R^x_{i,j}+u^x_{i,j})^2}{r^3_{i,j}}\\
		\partial n^x_{i,j}/ \partial u^y_{i,j}&= \partial( \frac{R^x_{i,j}+u^x_{i,j}}{r_{i,j}})/ \partial u^y_{i,j} = -\frac{(R^x_{i,j}+u^x_{i,j})(R^y_{i,j}+u^y_{i,j})}{r^3_{i,j}}\\
		\partial n^y_{i,j}/ \partial u^x_{i,j}&=\partial( \frac{R^y_{i,j}+u^y_{i,j}}{r_{i,j}})/ \partial u^x_{i,j} = -\frac{(R^y_{i,j}+u^y_{i,j})(R^x_{i,j}+u^x_{i,j})}{r^3_{i,j}}\\
		\partial n^y_{i,j}/ \partial u^y_{i,j}&=\partial( \frac{R^y_{i,j}+u^y_{i,j}}{r_{i,j}})/ \partial u^y_{i,j} = \frac{1}{r_{i,j}}-\frac{(R^y_{i,j}+u^y_{i,j})^2}{r^3_{i,j}}
		\end{split}
		\end{align}
		The above expressions can be written in a compact form using tensor notation $\nabla _{\mathbf{u}}\mathbf{n}_{i,j} = \frac{\mathbf{I}}{r_{i,j}}-\frac{\left(\mathbf{R}_{i,j}+\mathbf{u}_{i,j}\right)\otimes \left(\mathbf{R}_{i,j}+\mathbf{u}_{i,j}\right)}{r_{i,j}^3}$, which is adopted in  Eq.~\ref{S2}.
		
		\subsection{Preliminary stability analysis of the magnetic particle oscillator}
		As discussed in the example of the magnetic particle oscillator, instabilities may occur for certain configurations. For instance, in the case of $\theta=\pi/2$ shown in FIG.2 (d), the system approaches a 1D configuration, and we will show that its dynamical behavior becomes unstable in the horizontal direction. For the choice of parameters used in the article ($M=1$, $L_0=1$, $\gamma=10^4$), the complete stiffness matrix $\mathbf{D}$ is calculated as 
		\begin{align}\label{stifness_matrix_1D}
		\begin{split}
		\mathbf{D}&=\mathbf{D}_0 +\mathbf{D}^*\\
		&=\begin{bmatrix} 0 & 0 \\ 0 & 120000 \end{bmatrix}+\begin{bmatrix} -30000 & 0 \\ 0 & 0 \end{bmatrix}\\
		&=\begin{bmatrix} -30000 & 0 \\ 0 & 120000 \end{bmatrix}
		\end{split}
		\end{align} 
		where we notice that the first diagonal component is negative. From basic structural dynamics, a negative stiffness component is often related to dynamical instabilities. In this case, the instabilities occur when the resonator is perturbed in the horizontal direction, which is in response to the fact that there is no horizontal resistance provided in the system. To prevent the occurrence of instabilities in this configuration, the horizontal motion of the resonator is constrained in the simulation. This result provides additional evidence that our analytical model is capable of capturing the complete dynamical behavior of the oscillating particle system.
		
		\subsection{A modified analytical model for the lattice prototype}
\begin{figure} [!htb]
			\centering
			\includegraphics[scale=0.13]{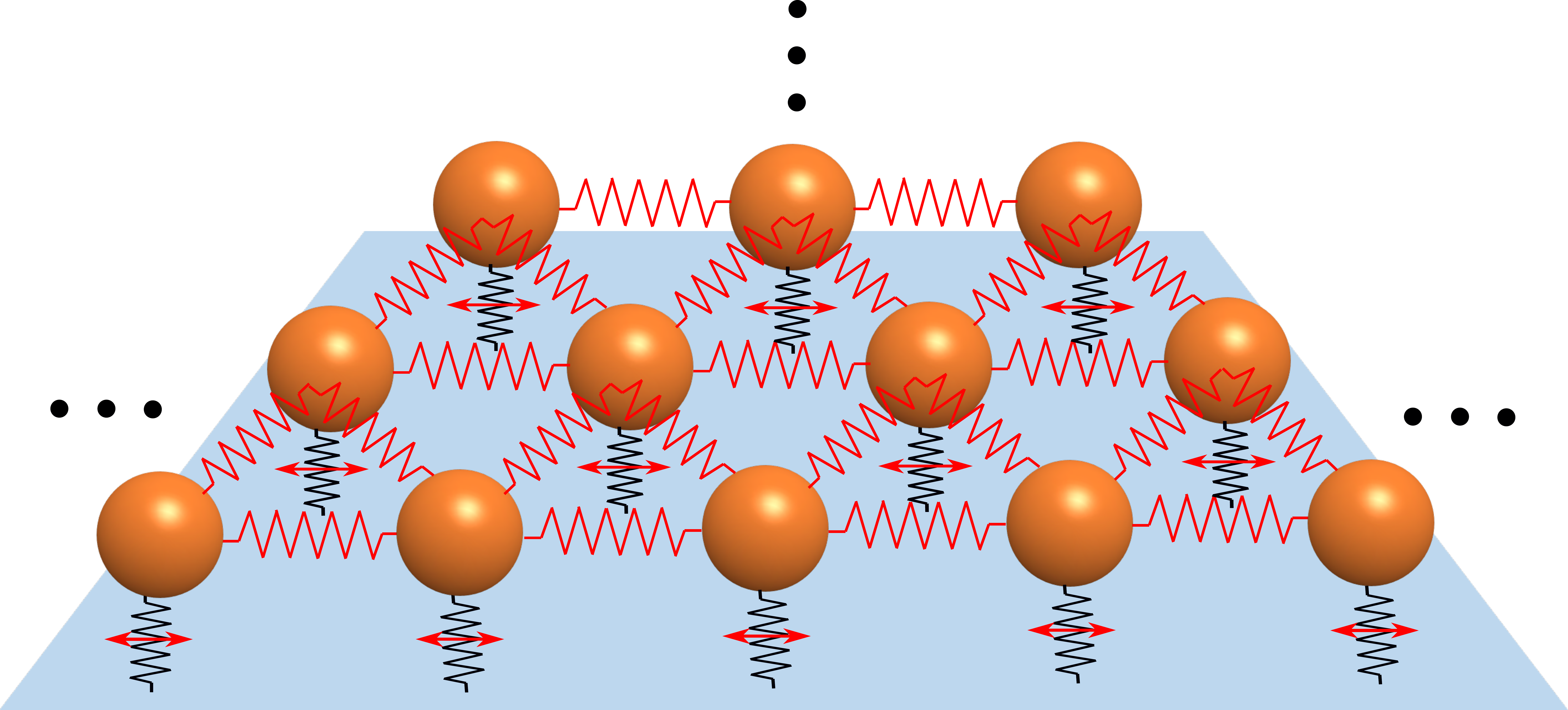}
			\caption{Equivalent spring-mass model of the lattice specimen, featuring an elastic foundation with in-plane spring constant.}
			\label{SFig1}
		\end{figure}

		To provide guidelines for the experiment, we propose a modified analytical model to approximately capture the dynamics of the lattice specimen, as shown in Fig.~\ref{SFig1}. The magnets are modeled as point masses connected by springs featuring repulsive interaction law $f(r)$. The supporting beams are treated as an elastic foundation with in-plane equivalent spring constant
		\begin{equation}
		k_{eq}=\frac{3EI}{H^3}	
		\end{equation}
		typical of cantilever beams, where $E$ is the Young's modulus, $I=\pi R^4/4$ is the second moment of area of the beam's circular cross-section, and $R$ and $H$ are the radius and height of the beam.
		With the elastic foundation incorporated to the analytical model introduced in Sec.~IV~A, the governing equation for this modified system becomes
		\begin{equation}\label{Governing_equation_modified}
		\mathbf{M}\mathbf{\ddot{u}}_{i,j}+\sum_{l=1}^{3}\left\lbrace \left[  f_r(L_0)\mathbf{e}_l\otimes    \mathbf{e}_l  + \frac{f(L_0)}{L_0}\left( \mathbf{I}-\mathbf{e}_l \otimes \mathbf{e}_l\right) 
		\right]\Delta \mathbf{u}_l \right\rbrace + \mathbf{K}_f\mathbf{u}_{i,j}=\mathbf{0}
		\end{equation}
		where the additional stiffness term $\mathbf{K}_f=\begin{bmatrix} k_{eq} & 0 \\ 0 & k_{eq} \end{bmatrix}$ accounts for the resistance effect of the elastic foundation, and the rest quantities are defined in the main text.
		
\begin{figure} [!htb]
			\centering
			\includegraphics[scale=0.15]{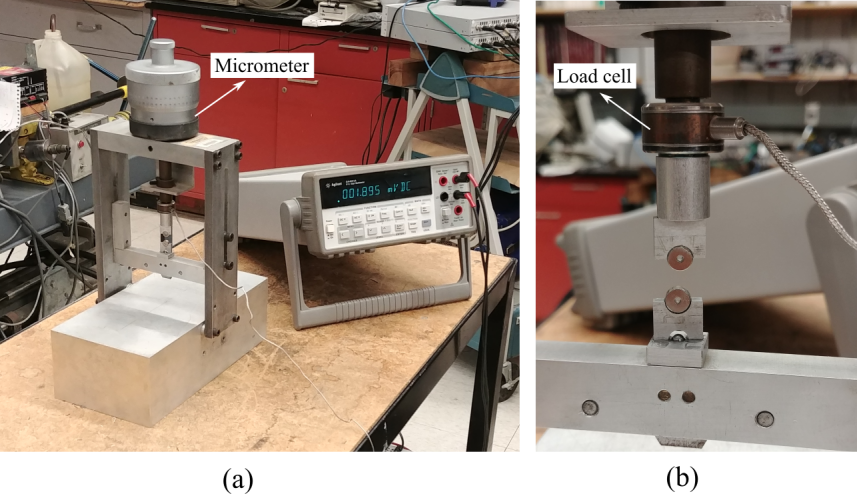}
			\caption{Experimental characterization of the repulsive force between magnets. (a) Testing apparatus. (b) Detail of the magnets position.}
			\label{SFig2}
		\end{figure}

\begin{figure} [!htb]
			\centering
			\includegraphics[scale=0.8]{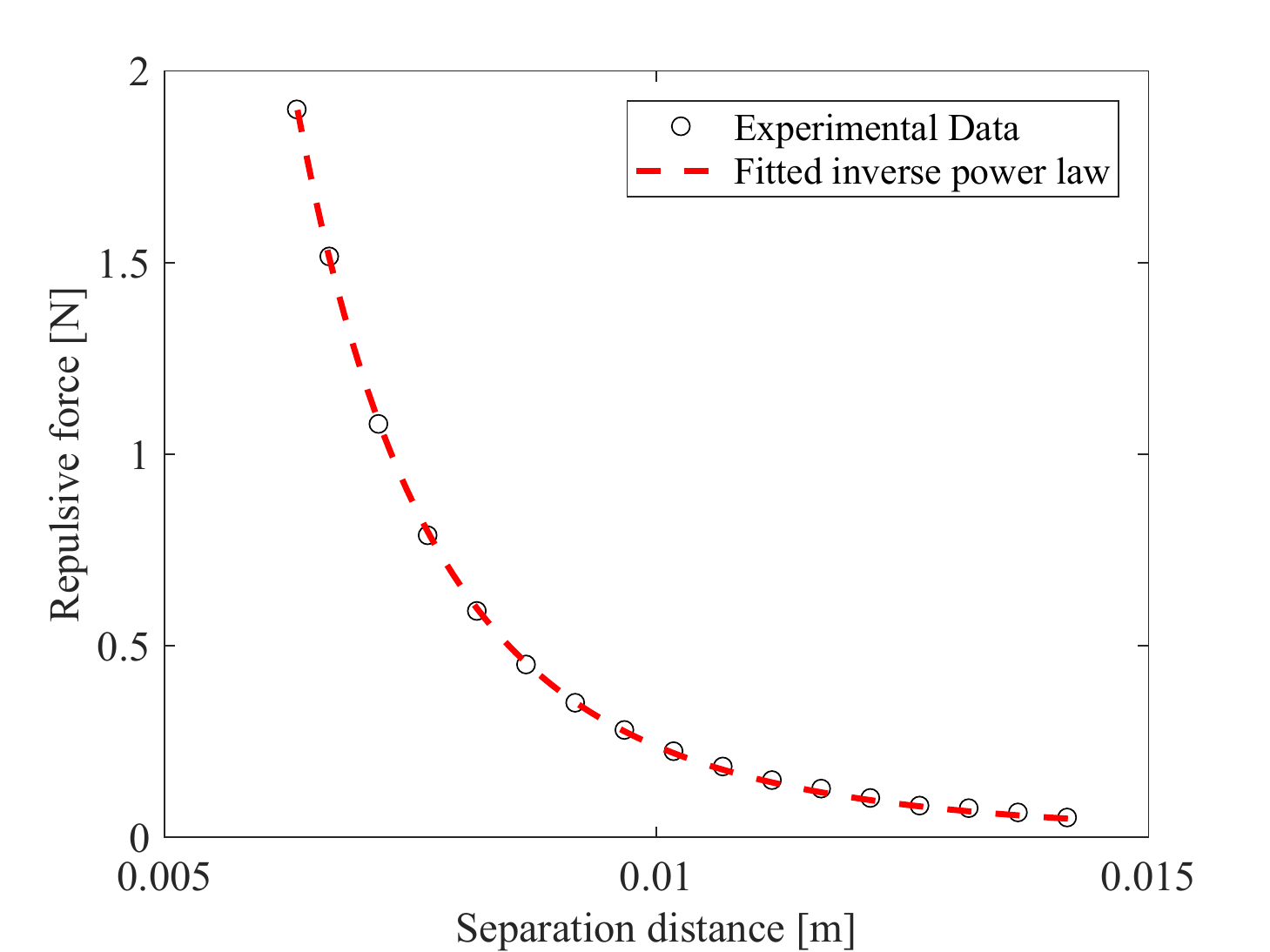}
			\caption{Experimental data and fitted force-displacement curve for two magnets with repulsive interaction.}
			\label{SFig3}
		\end{figure}

		Considering Bloch conditions (Eq.~16), the wavenumber-dependent stiffness matrix for the modified analytical model can be written as
		\begin{equation}
		\mathbf{D}'=\mathbf{D}+\mathbf{K}_f	
		\end{equation}
		where $\mathbf{D}=\mathbf{D}_0+\mathbf{D}^*$ and its expression is given in Eq.~18. The dispersion relation can be obtained by solving the corresponding eigenvalue problem with the modified stiffness matrix $\mathbf{D}'$, which requires determining $f(L_0)$ and $f_r(L_0)$ from the repulsive interaction law between pairs of magnets. To this end, we conduct a static test on two magnets. The experimental setup is shown in Fig.~\ref{SFig2} (a). The two magnets are mounted on appropriately machined holder such that their sides can progressively moved against each other. A micrometer is used to control by small amounts the distance between the two magnets, and a highly sensitive load cell (shown in Fig.~\ref{SFig2} (b)) is attached to the top magnet grip to measure the repulsive force. The measured force values are plotted as circles in Fig.~\ref{SFig3}, and fitted by an inverse power law $f(r)=b r^{-a}$ using ``lsqcurvefit" in Matlab (based on the method of nonlinear least squares). We determine $a = 4.5824$ and $b = 1.6209 \times 10^{-10}$  (the dashed red line in Fig.~\ref{SFig3}). 
		
\begin{figure} [!htb]
			\centering
			\includegraphics[scale=0.8]{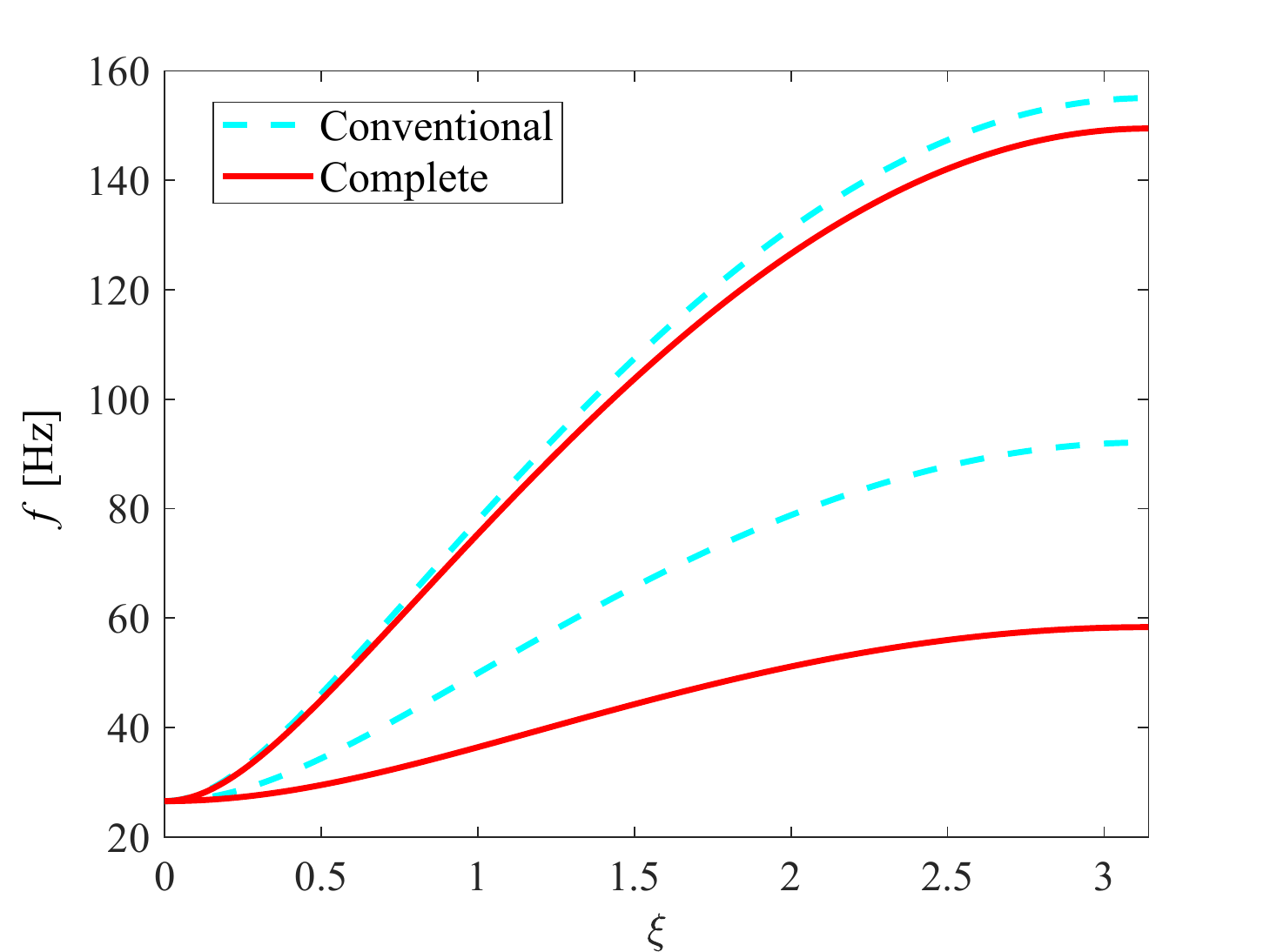}
			\caption{Comparison between dispersion relation of the modified analytical model and that of the conventional (reference) model, showing large mode-selective dispersion shifts.}
			\label{SFig4}
		\end{figure}

		The other parameters necessary for the calculation of the eigenvalue problem are listed below: the mass of the magnet $M=7.07  \times 10^{-4}$ kg, the Young's modulus of Aluminum $E=71$ Gpa, the diameter and height of the beam $R=7.9375 \times 10^{-4}$ m and  $H= 0.15$ m, and the distance between two magnets (at rest) in the lattice $L_0=0.01$ m. Finally, we determine from the modified analytical model the dispersion relation for wavevectors along the $\mathrm{\Gamma}$-$\mathrm{M}$ direction, which is plotted in Fig.~\ref{SFig4}. For comparison, the reference dispersion relation, conventionally obtained for the system with stiffness matrix $\mathbf{D}_0+\mathbf{K}_f$, is superimposed as dashed blue lines.

\end{document}